# Integrated Assessment Modeling of Korea's 2050 Carbon Neutrality Technology Pathways

Hanwoong Kim[1], Haewon McJeon[2], Dawoon Jung[1], Hanju Lee[1], Candelaria Bergero[2], Jiyong Eom[1]*

[1]College of Business, Korea Advanced Institute of Science and Technology (KAIST), Republic of Korea
[2]Joint Global Change Research Institute, Pacific Northwest National Laboratory (PNNL), USA
*Corresponding author: eomjiyong@kaist.ac.kr

## Abstract

This integrated assessment modeling research analyzes what Korea's 2050 carbon neutrality would require for the national energy system and the role of the power sector concerning the availability of critical mitigation technologies. Our scenario-based assessments show that Korea's current policy falls short of what the nation's carbon-neutrality ambition would require. Across all technology scenarios examined in this study, extensive and rapid energy system transition is imperative, requiring the large-scale deployment of renewables and carbon capture & storage (CCS) early on and negative emission technologies (NETs) by the mid-century. Importantly, rapid decarbonization of the power sector that goes with rapid electrification of end-uses seems to be a robust national decarbonization strategy. Furthermore, we contextualize our net-zero scenario results using policy costs, requirements for natural resources, and the expansion rate of zero-carbon technologies. We find that the availability of nuclear power lowers the required expansion rate of renewables and CCS, alleviating any stress on terrestrial and geological systems. By contrast, the limited availability of CCS without nuclear power necessarily demands a very high penetration of renewables and significantly high policy compliance costs, which would decrease the feasibility of achieving the carbon neutrality target.

# 1. Introduction

A number of countries, municipalities, and corporations have made carbon neutrality or net-zero commitments, beginning to take concrete steps to achieve these ambitious goals [1, 2]. Unless their leadership mobilizes global mitigation actions required to achieve global net-zero around 2050, the world may face significant economic losses due to climate change [3]. This is why the Paris Agreement set out the ambitious goal of keeping the global temperature rise to well below 2°C or 1.5°C [4, 5]. Rapid and concrete national mitigation strategies are needed to meet the ambitious climate goal. Against this backdrop, three East Asian economies with the combined GHG emissions accounting for 33% of the world's total made carbon neutrality pledges by 2050 (Japan and Korea) or 2060 (China) [6-8]. The declarations have drawn considerable attention for the countries' significance as a share of global Greenhouse Gas (GHG) and for their policy ambition.

In compliance with the Paris Agreement, Korea pledged as the Nationally Determined Contribution [9] to achieve a 24.4% reduction in GHG emissions by 2030 relative to the 2017 level. Ratcheting up the ambition, the President of Korea declared in October 2020 the nation's 2050 carbon neutrality strategy as a comprehensive and rapid transition plan towards a sustainable and low-carbon economy [6]. However, the prevailing view is that the Korea's current policy measures may be incompatible with its 2030 NDC target, as well as the 2050 carbon neutrality goal. [10-12]. Therefore, an objective assessment of Korea's 2050 carbon-neutrality scenarios that accounts for its energy policy and technology options is needed at this crucial moment with less than 30 years left to achieve the carbon neutrality.

Our study addresses the following three questions: (i) What does the 2050 carbon neutrality target mean for Korea's energy system? (ii) What is the role of the power sector for the 2050 carbon neutrality target vis-à-vis the other sectors, and how important is the availability of key mitigation technologies? (iii) What is the feasibility of the alternative carbon neutrality scenarios and tradeoffs between constraints associated with the technology deployment? We compare a reference scenario reflecting current national policies with three 2050 carbon-neutrality scenarios varying in the availability of critical mitigation technologies such as nuclear power, renewable energy and carbon capture and sequestration (CCS).

This paper, as the first academic study assessing Korea's 2050 carbon neutrality technology pathways, makes two primary contributions. First, it contributes to our understanding of the potential implications of Korea's carbon neutrality policy regarding the extent of energy transition and associated compliance costs and their interaction with the availability of effective but controversial mitigation options such as nuclear power and CCS. Many integrated assessments have focused on a subset of these aspects in the context of energy system decarbonization at the national level. Previous studies suggest that the transition requires rapid deployment of various low- to zero-carbon energy technologies and rapid phase-out of fossil fuel-based infrastructures [13-19], the availability and viability of key technology options such as nuclear power, renewables, and CCS can help reduce policy compliance costs depending on the nature of the current energy system and policy timelines and stringency [10] and the decarbonization of the power sector coupled with the electrification of the end-use sectors can be an essential enabler of national decarbonization policies [20]. Our study distinguishes itself from these previous works, in that we first analyze 2050 carbon-neutrality scenarios for Korea using an integrated assessment modeling framework. The challenges faced by Korea are unique and worth academic investigation because the nation is pursuing a very rapid carbon-neutrality transition of the carbon-intensive energy system, which have been continuously expanding since 1948 when the country was established. Korea's carbon-neutrality case also set an example for rapidly industrializing nations around the world. Our study is also policy-relevant as the nation's resource potential for renewables and carbon sequestration is considered limited compared to its high knowledge capital for nuclear power, which is planned to be phased out over time. A representation of Korea's national policies and technology options within the long-term integrated assessment framework, the Global Change Assessment Model (GCAM), allowed us to explore the implications that the availability of critical mitigation options have for the decarbonization of Korea's entire energy system.

The second contribution of this paper is that it examines the feasibility tradeoffs between constraints to the deployment of key mitigation technologies underlying Korea's unique policy challenges. We show that currently considered national policy targets and technology options may risk facing feasibility issues. Techno-economic studies traditionally considered such feasibility factors as resource and technological potential and policy compliance costs to achieve particular emissions mitigation targets [10, 21]. More recent feasibility studies have begun to pay attention to public and societal acceptance [17, 22] and a more complete set of feasibility factors, such as techno-economic, socio-technical, and political aspects [23]. In light of Korea's historical experience of energy system development and resource potential, we assess the speed and scale of expanding low- and zero-carbon technologies and phasing out fossil fuels as required by the carbon-neutrality transition.

Our exercise suggests a dramatic low-carbon transition over the next three decades is necessary for Korea's carbon neutrality across all technology scenarios. As anticipated for the low-carbon transition of other economies, the power sector plays a pivotal role in Korea's low-carbon transition through the sector's decarbonization and associated electrification of individual end-use sectors. Coal power phases out by the 2030s, and in the longer term, negative emissions technologies (NETs) are deployed in a large proportion to offset hard-to-abate emissions from the economy. Our technology scenarios indicate that, although renewable power continues to scale up over the next three decades, the expansion rate shows wide variation depending on the nuclear power and CCS availability. The availability of nuclear power lowers the required rate of expansion of renewables and CCS, alleviating the stress on terrestrial and geological systems. However, the limited availability of CCS without nuclear power necessitates a very high penetration of renewables and significantly high policy compliance costs, which may hinder the feasibility of achieving the carbon neutrality target.

The remainder of the paper is structured as follows. Section 2 provides background for Korea's climate policy and energy system and literature on the nation's decarbonization. Section 3 briefly introduces the GCAM-KAIST integrated assessment model and scenarios explored in our paper. Section 4 discusses the results emerging from the scenarios, and concluding remarks are presented in Section 5.

# 2. Background

## *2.1. Korea's climate policy and energy system*

To respond to the call for GHG mitigation under the Paris Agreement, the Korean government has pledged a 24.4% reduction in GHG emissions by 2030 relative to the 2017 level for the Nationally Determined Contribution [9]. Participating in the increased ambition of many countries to achieve net-zero by around 2050, the President of Korea declared in October 2020 the 2050 carbon neutrality strategy as a comprehensive and rapid transition plan towards a sustainable and low-carbon economy [6]. Detailed policy instruments and technology development plans are to be announced in the future. However, the strong policy push for the decarbonization and carbon neutrality of the economy presents unique challenges for the three reasons.

First, the energy system transition plan calls for the rapid withdrawal of coal as well as the gradual phase-out of nuclear power. The two baseload power sources currently account for 64.6% of electricity generation and 47% of total installed capacity (KEPCO).[a] The 9th National Power Supply Plan (NPSP) made it clear that 30 out of the current 60 coal units will close by 2034 [24]. With the ratcheting of national climate policies and increasing environmental activism, the remaining units are also expected to be under-utilized or stranded before the mid-century. The NPSP also decided to discontinue any new construction of nuclear power, gradually decreasing its installed capacity from 23.3 GW to 19.4 GW by 2034, even though Korea has internationally recognized capabilities for advanced nuclear power design, construction, and operation [25]. Instead, the sector will secure 58.1 GW from LNG units (including the conversion of 24 coal units) and 77.8 GW from new and renewable sources by 2034. The pursuit of cleaner power sources with rapidly decreasing reliance on the two major baseload technologies would present a major challenge to the power system reliability unless significant investments are made in flexibility resources such as utility-scale energy storage systems.

---

[a] We complied annual statistics for power plants data presented in the KEPCO reports. They are available in the following link (in Korean):
https://home.kepco.co.kr/kepco/KO/ntcob/list.do?boardCd=BRD_000099&menuCd=FN05030103

Second, new and clean technologies requiring intensive land and geologic resources need to be rapidly scaled up to enable the carbon-neutral transition of the energy system. Although renewable energy capacity in Korea has increased steadily since 2005, the growth rate is far behind G20 countries [26]. By 2018, Korea had the lowest share of renewable energy in energy supply among IEA countries [27]. With this backdrop, the NPSP has set the ambitious deployment plan of more than tripling renewable capacity from 20.1 GW to 77.8 GW by 2034. The recent declaration of the 2050 carbon neutrality is likely to require an upward adjustment of the target. However, given that Korea has a relatively small landmass compared to its large population and economic output, scaling up renewables and CCS to a level consistent with the plans may be particularly challenging [28].

Third, Korea's end-use sectors, such as the industry, transportation, and buildings, have remained carbon-intensive, accounting for 58% of total GHG emissions (Ministry of Environment)[b]. Thus, end-use sectors should take a considerable part in the energy system transition. The rapid decarbonization of end-use sectors, however, seems challenging, particularly for the industry and transportation sectors. Manufacturing has been the main driver of Korean economic development and thus contributed significantly to national GHG emissions (37% in 2017) (Ministry of Environment, 2021)[c]. Although various price- and nonprice-based measures to promote low-carbon investments are underway, few viable technology options are available for highly carbon-intensive sectors in Korea, such as steel, cement, and petrochemicals. Transportation, particularly freight, is another hard-to-abate end-use sector, accounting for 13.5% of total GHG emissions (Ministry of Environment, 2021)[d]. Although the government has implemented fuel efficiency measures and accelerated the deployment of electric and hydrogen vehicles, the progress has been slow, with gasoline and diesel accounting for more than 97% of transportation's total energy consumption. The buildings sector accounts for about 7% of national emissions (Ministry of Environment, 2021)[e]. Investments in energy efficiency would help the decarbonization of the sector, with the rest from fuel switching from gas or oil to electricity.

---

[b]  We complied annual statistics for GHG emissions presented in the GIR reports. They are available in the following link (in Korean): https://www.index.go.kr/potal/main/EachDtlPageDetail.do?idx_cd=1464

[c]  We complied annual statistics for GHG emissions presented in the KEPCO reports. They are available in the following link (in Korean): https://www.index.go.kr/potal/main/EachDtlPageDetail.do?idx_cd=1464

## 2.2. Literature on the decarbonization of Korea and others

Except for limited academic or grey literature on energy system development towards a low-carbon economy, our study is the first one that explicitly investigates the role of critical mitigation technologies and associated feasibility considerations for the carbon-neutrality transition in Korea. Park et al. [13] explored alternative technology scenarios for the Korean power sector, demonstrating that renewable-based power systems can promote sustainable development, even though the discounted system costs are 20% higher than the case relying on nuclear and coal power. Hong et al. [14] studied the impact of alternative development scenarios on the energy system, conducing a comparative analysis of several response indicators, such as energy security, job creation, and GHG emissions. The study shows that increasing the share of renewable energy and decreasing energy demand can promote energy security and employment in the energy sector while reducing GHG emissions. However, these studies do not assess the transition of energy systems under the unprecedentedly stringent carbon-neutrality target. To the best of our knowledge, Jung and Park [10] is the only study on more stringent energy transition consistent with the ambition of the Paris agreement. The study assessed the feasibility of reducing national GHG emissions to achieve zero-emissions target by 2055, indicating that economy-wide carbon tax is necessary to enable a drastic change in the energy system. In terms of technology options, the study suggests that Korea needs to increase the share of nuclear and renewable energy to achieve more than an 80% reduction in GHG emissions by 2050. Yet, it does not evaluate various mitigation technology options and potential tradeoffs between technology deployment constraints under the stringent carbon neutrality target.

A relatively large amount of literature exists on the implications of carbon-neutrality policy for other countries or global energy systems. Clarke et al. [29] provides insights into the global development of zero- and low-carbon energy supply, including renewables, nuclear, and CCS technology. This study indicates that achieving the stringent 2°C climate stabilization target entails the large-scale deployment after 2050 of negative emissions technologies (NETs), such as bioenergy with CCS (BECSS), land use, land-use change, and forestry (LULUCF), and direct air capture (DAC), to offset residual emissions from hard-to-abate sectors, such as industry and aviation [30, 31]. Fragkos et al. [11] also shows that the national low-emission pathways of eleven world regions are consistent with what is required to limit global temperature rise to well below 2°C. For the authors, emissions reductions are achieved by deploying various mitigation technologies and measures, such as renewables, nuclear power, CCS technology, energy efficiency, and end-use electrification. Their study also suggests that NETs will be necessary around the mid-century to offset residual emissions or become carbon negative. The literature collectively suggests that the carbon-neutrality transition of energy systems will require a significant shift in investment towards low- and zero-carbon technologies and NETs.

---

[d]  We complied annual statistics for GHG emissions presented in the KEPCO reports. They are available in the following link (in Korean): https://www.index.go.kr/potal/main/EachDtlPageDetail.do?idx_cd=1464

[e]  We complied annual statistics for GHG emissions presented in the KEPCO reports. They are available in the following link (in Korean): https://www.index.go.kr/potal/main/EachDtlPageDetail.do?idx_cd=1464

Among country-level net-zero transition studies are China's deep decarbonization assessment. Zhao et al. [32] analyzed China's national emissions pathways to achieve its NDC and 2060 net-zero target. The authors' conclusion is that during 2020-2050, the power sector would bear the largest emissions reduction burden, followed by the industry, transportation, and building sectors, with a significantly larger share of electricity production coming from renewables and nuclear energy. Another study on China's net-zero target shows that negative emissions play a prominent role, offsetting 3 GtCO2 emissions per year from remaining emitting sectors, such as freight transportation and heavy industry [33]. The negative emissions amount includes up to a 1.6 GtCO2 per year contribution from DAC. When it comes to Japan's NDC and 2050 net-zero target, Oshiro et al. [21] indicates that the implementation of the NDC can accelerate a transition from its baseline pathway, thanks to improved energy efficiency and the decarbonization of electricity. However, the longer-term 2050 net-zero requires a rapid, extensive transformation of the energy system post-2030, involving large-scale renewables and CCS deployment, energy efficiency, and electrification. Kato and Kurosawa [34] examine in more depth the role of NETs, BECCS, and DAC in particular, suggesting that a large-scale deployment of NETs is imperative to achieve Japan's net-zero vision as part of its long-term strategy. In a similar vein, Bergero et al. [15] have assessed NDC and net-zero pathways for Taiwan, showing that the net-zero target would require an aggressive electrification of end-use sectors, the rapid expansion of renewables, and increased efficiency in end-use sectors. It also suggested that NETs become critical to offset harder-to-decarbonize sectors, such as heavy transportation and industry.

There also are country-level studies on timing and uncertainties associated with net-zero transitions. Williams et al. [35] guides carbon neutrality policy with concrete near-term priorities. The authors find several feasible options to supply low-carbon fuels for non-electrifiable industrial, freight, and aviation end-uses, which would be required on a large scale only after 2035. The required near-term actions remain primarily similar across all pathways: expanding renewable capacity, decommissioning coal-fired power stations, maintaining existing gas-fired capacity, and increasing sales of electric vehicles and heat pumps. Chaturvedi et al. [36] shows how scenario-based uncertainty assessment for India can inform NDC and longer-term decarbonization. The authors identify renewable integration costs and decarbonization of industrial energy use as key challenges, calling for reforms of the electricity sector to address these challenges.

The above literature commonly provides four key insights into requirements for achieving carbon-neutrality or net-zero. First, the rapid penetration of renewables is needed over the next three decades at national and global levels. Second, despite uncertainties surrounding CCS technology, it can accelerate the power sector's decarbonization reliably. Third, decarbonizing the power sector coupled with electrifying end-uses can be an effective national decarbonization strategy. Lastly, in the longer term, NETs would be needed to come online in a large proportion to offset hard-to-abate emissions from the economy. The current paper investigates required energy system development under different technology availability assumptions and how such requirements might constrain each other as Korea pursues its 2050 carbon-neutrality target.

## 3. SCENARIOS AND METHOS

### 3.1. SCENARIOS

We have developed four alternative scenarios to explore a possible range of future pathways relevant to understanding the energy system requirements for Korea's 2050 carbon neutrality target. These scenarios, which are summarized in Table 1, are explained below in order:

The current policy scenario (CurPol) reflects Korea's current and planned energy and climate policy measures. It can thus be called the reference policy scenario. The scenario's electricity generation profiles follow the 9th National Power Supply Plan [24] and the 5th Renewable Energy Basic Plan [37]. The forecasts of total energy demand and energy consumption of individual end-uses reflect the 3rd Basic Energy Plan [38]. Future technology deployment and associated GHG emissions in the transportation sector, are consistent with the 4th Eco-friendly Vehicle Basic Plan [39], the Corporate Average Fuel Economy, and the GHG Emissions Management Program [40].

The CurPol scenario is compared against the three carbon-neutrality scenarios with varying assumptions on critical mitigation technologies. First, the 2050 net-zero scenario (NZ2050) presupposes policy instruments employed in the CurPol scenario but implements additional climate policies to achieve net-zero GHG emissions by 2050. Specifically, the nation's annual GHG emissions are assumed to decrease linearly until they reach net-zero in 2050. The emissions pathway represents what it takes to make immediate, progressively intensified mitigation efforts to net zero emissions by 2050. Second, the nuclear-assisted net-zero scenario (NZ2050_Nuc) shares the same assumption as the NZ2050 scenario, except that the expansion of nuclear power is allowed to deploy based on its economics. Third, the CCS-constrained net-zero scenario (NZ2050_limCCS) also shares the same assumptions as the NZ2050 scenario, except that CCS deployment is restricted due to assumed higher capital costs.

## 3.2. METHODS

This research uses the Global Change Analysis Model (GCAM) to assess different net-zero trajectories for Korea. GCAM was developed by U.S.-based PNNL/JGCRI and is a representative energy-economy-environment assessment model which has been routinely used in major climate policy evaluation studies, including IPCC reports system (publicly available at https://github.com/JGCRI/gcam-core). GCAM is a partial equilibrium model [41] and an IAM [42]. It is also one of the five representative models used for the development of Representative Concentration Pathways (RCP) and Shared Socioeconomic Pathways (SSP). GCAM is an IAM that links macroeconomics to energy systems, land use, water, and climate systems, and can be used to develop a consistent scenario that allows an integrated assessment of the impact of energy and climate policies on each energy and land system from various perspectives.

GCAM v.5.2 divides the world into 32 different regions (Korea is classified as one single region out of the 32 regions), and classified 235 river basins, 384 subregions, and 24 GHGs. All the regions are linked through international trade in energy commodities, agricultural, and other goods. At the macro level, the scale of economic activity is given by population and labor productivity that determine economic output. GCAM's energy system reflects competition among fuels and technologies from primary energy production, energy transportation, and end-use energy consumption by region and transaction of energy goods between regions. Energy transformation sectors convert primary energy resources into secondary energy carriers, which are ultimately converted into goods and services, demanded by end-use sectors. In terms of technology competition, multiple technologies compete for market share implementing logit choice model [41]. The cost of a technology in any period depends on two key exogenous input parameters: the non- energy cost and the efficiency of energy transformation as well as the prices of the fuels it consumes. The efficiency of a technology determines the amount of fuel required to produce each unit of output. The prices of fuels are calculated endogenously in each time period based on supplies, demands, and resource depletion. The model uses its base year of 2010 and runs from 2015 until 2100 in five-year time steps.

There are three end-use sectors that consume final energy, namely buildings, industry, and transportation. The building sector is divided into residential and commercial buildings, for which three services are modeled: heating, cooling, and others. Demand for each energy service grows as a function of income and service prices. The industrial sector includes all industrial activities except refining, cement, and fertilizers, which are modeled separately in GCAM. The industrial sector consumes energy and feedstocks based on cost competition. Demand for industrial services responds to income growth and fuel prices. The third end-use sector is the transportation sector, which is broadly divided into four sectors based on the mode of travel. Energy demand in the transportation sector is modeled for passenger transport, freight transport, and international transport, with demand for each service determined by per capita GDP and population.

GCAM-KAIST 1.0 is an extended version of the default GCAM v5.2 developed to address our research questions. GCAM-KAIST 1.0 enhances GCAM v5.2 to represent Korea's present and planned policies and technologies and thus to provide more robust policy implications. The power sector in our extended model reflects the power plant deployment plan and electricity generation outlook through 2034 specified in the 9th Basic Plan for Electricity Supply and Demand [24] and the 5th Renewable Energy Basic Plan [37]. Concerning end-use sectors, the industry and building sectors deploy measures to reproduce the target energy demand specified in the 3rd Basic Energy Plan [38]. The transport sector reflects the 4th Eco-friendly Vehicle Basic Plan [39], and GHG emissions projections for passenger vehicles are aligned with the management goals under the Vehicle Average Fuel Economy and GHG Emissions Management Program [40]. For more details, see Supplementary Material (Section 2).

# 4. Results

## *4.1. Mitigation of GHG Emissions for 2050 Carbon Neutrality*

Based on our analysis, GHG emissions of the current policy scenario ("CurPol", 2% reduction relative to the 2017 level by 2030) fall short not only of Korea's NDC target[f] (24.4% reduction by 2030 or, equivalently, 536 MtCO2e) but also of what the nation's 2050 carbon-neutrality ambition would require (Figure 1). Both the current policy scenario and the nation's current NDC target seem to present significant gaps with the straight path towards 2050 carbon neutrality (NZ2050 scenario), which requires a 31% emission reduction by 2030.

Cost-effective achievement of the 2050 carbon neutrality target would determine the sectoral mitigation burden and its development over time (Figure 2). Our analysis suggests that the most significant contribution to the carbon neutrality should come from the power and industry sectors, followed by the transportation and buildings sectors. Yet, there is a minor difference in sectoral GHG mitigation burden under the three different net-zero scenarios. As the power sector decarbonization comes cheaper with NZ2050_Nuc than NZ2050, the sector's mitigation burden is alleviated in NZ2050_Nuc and, thanks to reduced electricity prices, and passed onto end-use sectors, particularly the industrial sector. By contrast, as decarbonization is more expensive with NZ2050_limCCS than NZ2050, the mitigation burden of the power sector is not easily shared with end-use sectors in NZ2050_limCCS due to increased electricity prices (See Figure S3 in the Appendix).

Our results also suggest that NETs are critical to achieve 2050 carbon neutrality, regardless of whether critical mitigation options (renewables, CCS, and nuclear) become economically viable (Figure 3). However, the degree and manner in which NETs are introduced are different across the three carbon-neutrality scenarios. Compared to the NZ2050 case where a significant amount of residual emissions is offset by a considerable amount of negative emissions (35% by DAC 25% by biopower with CCS, 40% by LULUCF) by 2050, the cases of NZ2050_Nuc and NZ2050_limCCS generate less residual emissions and thereby less negative emissions by the mid-century. Lower residual emissions in NZ2050_Nuc (and associated lower negative emissions) than the NZ2050 case originate from faster electrification of end-use sectors due to lower electricity prices presented by the nuclear-assisted scenario. In contrast, lower residual emissions in NZ2050_limCCS (and associated lower negative emissions) come at the expense of economy-wide reductions in energy consumption by the mid-century as limited CCS opportunities necessarily put upward pressure on carbon prices.

---

[f] The existing NDC has been recently replaced by a new, more stringent target that requires a 40% GHG reduction relative to the 2018 level. This revision will go through a deliberation process and will be submitted to UNFCC in December 2021.

### 4.2. Requirements for Energy System Transition for 2050 Carbon Neutrality

The rapid reduction in GHG emissions would require a drastic change in the national energy system, which we discuss below from the top down. In terms of primary energy consumption (Figure 3), we find a decrease in oil and coal but an increase in natural gas, as well as a rapid expansion of renewable sources such as solar, wind, and biomass (about an eight-time rise over the next three decades), in the case without the carbon neutrality policy between 2010 and 2050 (CurPol scenario). However, in the carbon-neutrality scenarios, the transition from fossil fuels toward renewables and CCS dramatically accelerates. Regarding the extent of expansion in renewables over the next three decades, the NZ2050 scenario requires a 27-time increase in solar and a 38-time increase in wind energy. The NZ2050_Nuc presents relatively slow 18- and 27-time additions in solar and wind, respectively, thanks to the increased deployment of nuclear power. Finally, the NZ2050_limCCS scenario requires most rapid 36- and 46-time increases as CCS-based mitigation options become constrained.

The rapid transition in the primary energy system is enabled by the dramatic decarbonization of the power sector (Figure 4). Our scenario results on the power sector provide four key findings. First, even if the current policy stance continues (CurPol), the power sector experiences rapid and significant changes in the energy mix over the next three decades. While the share of coal power decreases to 13% by 2035, the renewables and gas power shares increase to 23% and 40%, respectively, which is consistent with the nation's NPSP. By 2050, the shares of coal, renewables, and gas power split even further, reaching 0%, 45%, and 40%, respectively. These dramatic changes in combination produce a 72% decrease in the carbon intensity of power generation between 2020-2050.

Second, coal phase-out seems critical (Figure 4). Coal units are expected to phase out during the 2040s under the current policy regime (CurPol), whereas the coal phase-out occurs even earlier in the 2030s under the carbon neutrality policy, regardless of the availability of critical mitigation technologies. However, this does not mean the end of the fossil fuel era in Korea. Instead, the power system increases its reliance on gas power to fulfill its requirement for flexibility resources until it peaks in the 2030s and declines to be replaced by gas power installed with CCS and renewables equipped with energy storage systems (ESS).

Third, the carbon neutrality policy leads to a much more rapid, significant decarbonization than the CurPol case (Figure 5). In particular, the expansion of solar and wind stands out under the carbon neutrality policy, accounting for 65%, 43%, and 80% of total electricity generation by 2050 in the NZ2050_Nuc, NZ2050, NZ2050_limCCS cases, respectively (Figure 4). The NZ2050_Nuc case, however, significantly alleviates burdens on renewables with the rapid scale-up of nuclear power, accounting for 42% of total electricity generation by 2050. Regardless of technology scenarios, the carbon intensity of power generation decreases to zero before 2050 and turns negative, indicating that the sector starts delivering negative emissions from biopower-CCS during the 2040s. Although our reference carbon-neutrality scenario (NZ2050) tends to decarbonize slightly faster due to its higher reliance on biopower-CCS than nuclear-assisted (NZ2050_Nuc) or CCS-constrained scenarios (NZ2050_limCCS), their carbon intensities look broadly similar and rapidly decreasing.

Fourth, unlike the CurPol scenario, where the growth of the power sector levels off in the 2030s, all three net-zero scenarios indicate continued, substantial scale-up of the sector through 2050 (Figure 5). Over the next three decades, the electricity production increases by about 75% in our reference NZ2050 scenario, whereas it increases by about 85% in the nuclear-assisted scenario (NZ2050_Nuc) and the CCS-constrained scenario (NZ2050_limCCS). The faster electrification in the NZ2050_Nuc and NZ2050_limCCS cases occurs for different reasons. The former is attributable to lower electricity prices, whereas the latter is because the power sector's limited CCS opportunities force end-use sectors to accommodate relatively clean electricity ahead of schedule.

The rapid decarbonization of the power sector vis-a-vis its substantial scale-up over the next three decades presents considerable carbon mitigation opportunities from increasing the reliance of end-use sectors (i.e., industry, transportation, and buildings) on electricity, which is referred to as electrification. Our results indicate that end-use sectors will continue to experience electrification, with the rate much higher in the net-zero scenarios compared to the current policy scenario (Figure 6). The much faster electrification in the carbon neutrality cases compared to the CurPol scenario is due to the presence of rapidly rising carbon prices, which makes fossil fuels increasingly more expensive than electricity that rapidly decarbonizes. Interestingly, the three carbon-neutrality scenarios share a very similar electrification rate. Electrification appears to be the most effective carbon mitigation strategy compared to other options, irrespective of the availability of critical mitigation technologies.

The question arises of what these energy system transitions would require for individual end-use sectors. Concerning the transportation sector, its energy consumption continues to decrease in all scenarios because of the saturation of vehicle ownership, improvements in fuel economy, and, most importantly, the sizeable penetration of high-efficiency clean vehicle modes, such as electric vehicles (EV) and hydrogen vehicles. However, the sector's technology mix is very different depending upon whether the 2050 carbon neutrality is pursued or not. Under the current policy regime, conventional internal combustion engines (ICE), despite their share being steadily eroded by EVs and hydrogen and biofuel vehicles, account for about 70% of energy consumption by 2050. With the carbon-neutrality policy in place, however, the share of ICE rapidly declines to about half by 2050, with the remainder given away to the cleaner technologies (i.e., EVs, hydrogen, biofuel, and natural gas vehicles) (Figure 7).

The structural shift in the transportation sector is more evident when expressed by the demand for passenger service (passenger-km) and freight service (tonne-km), as opposed to energy consumption (Figures 8). In the absence of the carbon-neutrality policy, the cleaner technologies expand their passenger and freight service shares to about 50% and 30%, respectively. When the carbon-neutrality policy is actively pursued, the shares of the cleaner technologies increase to about 70% and 75%. Moreover, the progress in the shares of transportation technologies is almost the same across the three net-zero scenarios. Among the cleaner technologies, the penetration of EVs is most remarkable and is the theme that holds regardless of the scenarios. The pursuit of carbon neutrality makes the role of electrification even more significant, increasing the share of EVs to more than half in both passenger and freight services.

The buildings sector distinguishes itself from other sectors in terms of its required transition (Figure 9). Even without the carbon-neutrality policy, electrification progresses rapidly over the next three decades, reaching about 65% of building energy consumption by 2050. This trend has to do with the electrification of heating and cooking services and increasing plug-load services associated with the assumed income increase. With the pursuit of 2050 carbon neutrality, electrification occurs even more quickly to about 85% by 2050 at the expense of decreased reliance on gas consumption.

The carbon-neutrality policy brings the largest changes to the industry sector (Figure 10). Under the CurPol scenario, the share of fossil fuels (coal, oil, and gas) remains nearly unchanged over the next three decades, despite the sector slowly decrease in coal. However, once the carbon-neutrality policy is pursued, coal rapidly phases-out, virtually vanishing by 2050, regardless of the availability of critical mitigation technologies. The main driver of this policy-induced transition is the electrification of industrial services that have so far relied mainly on coal and oil. By 2050, electricity accounts for more than half of the sector's energy consumption except feedstocks in all carbon-neutrality scenarios. It also is worth noting the steadily increasing use of hydrogen, which becomes the second-largest fuel source in the industry sector by the mid-century.

### 4.3. Feasibility of 2050 Carbon Neutrality

We now assess the feasibility and tradeoffs of the different carbon neutrality scenarios, including constraints associated with technology deployment. We do this by evaluating implied policy costs, land requirements, resource potential, and technology deployment rates.

When it comes to the cost of implementing 2050 carbon-neutrality, our reference carbon neutrality case (NZ2050) incurs about 2.4 percent of GDP by 2050. When the expansion of nuclear technology is allowed to deploy on an economic basis (NZ2050_Nuc), policy cost decreases over the next three decades, incurring about 1.7 percent of GDP by 2050. However, when CCS is constrained (NZ2050_limCCS), the transition becomes most costly, reaching 2.7 percent by 2050 (Figure 11). The cost assessment suggests that keeping the options of critical mitigation technologies wide open makes the ambitious carbon neutrality transition more economically feasible.

Careful consideration should also be given to the extent of natural resources required by the carbon-neutrality scenarios compared to the nation's resource endowment (Figure 12). For all scenarios examined, the required installed capacity of solar PV is well below the known technical potential of 973 GW, and the requirement for onshore and offshore wind power is also less than the technical potential of 352 GW and 387 GW, respectively [43].

However, some controversy can be expected regarding the availability of $CO_2$ storage reservoirs to fulfill the demand for the carbon-neutrality transition (Figure 13). Despite immense uncertainty regarding the available amount of $CO_2$ storage, one authoritative source reports that the total available $CO_2$ storage amount is estimated to be 1 $GtCO_2$ domestically, with the premise to explore additional 1 $GtCO_2$-size reservoirs abroad [6]. According to our analysis, the demand for $CO_2$ storage increases exponentially over the next three decades regardless of the three carbon-neutrality scenarios (Figure 13). The NZ2050 case requires the greatest cumulative storage amount of 1.9 $GtCO_2$ between 2015 and 2050 and is thus likely to be the first to hit the resource availability constraint. When the restrictions on nuclear construction are lifted (NZ2050_Nuc), the cumulative storage amount decreases by about 0.5 $GtCO_2$, relieving burdens on exploring new storage reservoirs. That is, decreasing reliance on nuclear power while pursuing 2050 carbon neutrality would require a much larger amount of CCS storage, which may exceed the nation's resource potential and deteriorate the feasibility of the carbon neutrality policy. The opposite of the NZ2050 case is the CCS-constrained case (NZ2050_limCCS), in which the cumulative storage amount is restricted to 0.6 $GtCO_2$ at the expense of the highest policy cost among the alternative scenarios.

Attention should also be paid to land area requirements, given that Korea has a relatively small landmass compared to its large population and considerable economic development. Although all three carbon-neutrality scenarios require substantial land areas for solar PV installation, the nuclear-assisted scenario (NZ2050_Nuc) seems most feasible in terms of land demand. The nuclear-assisted scenario (NZ2050_Nuc) requires roughly 30% less land for solar PV than the reference carbon-neutrality scenario (NZ2050) and about half less land than the CCS-constrained scenario (NZ2050_limCCS). However, the nuclear-assisted case (NZ2050_Nuc), would claim additional, geologically stable lands for siting increased nuclear facilities, which is about 10% of the size of Seoul by 2050 [44].[g] The most land intense scenario is the NZ2050_limCCS case, which deploys about 400GW of solar PV by 2050, and would require roughly 4.5 times the area of Seoul [45].[h] That is, securing enough $CO_2$ storage reservoirs is imperative to lessen the stress on land use.

Another possible criterion for feasibility assessment is to see if the expansion rates of major low-carbon technologies implied by the scenarios are consistent with the domestic or global technological capability as expressed by historical technology deployment experiences. The idea is that future development is reflective of the aggregate socio-technical mechanisms that have shaped the historical precedents [23]. Figure 12 shows that the required expansion rate of wind power over the next three decades remains not much different from Korea's historical experience and is largely insensitive to the availability of critical mitigation technologies.[i] However, the required expansion rate of solar PV far exceeds what the nation has experienced, particularly in the CCS-constrained case (NZ2050_limCCS). The nuclear-assisted scenario (NZ2050_Nuc) helps alleviate the urgency of solar PV expansion, promoting the feasibility of achieving the carbon-neutrality target. Note that the feasibility may not deteriorate in light of the expansion rate of nuclear power because even the most ambitious NZ2050_Nuc scenario presents a slower expansion of nuclear power than the historical experience (Figure 14).

---

[g] We assumed the unit land requirement of 0.745 $km^2$ per 1 GW nuclear power plant, following the National Assembly Budget Office (2017). The land size includes site area and sea area occupied by cooling water intake facility and hot water discharge facility.

[h] The unit land requirement is assumed to be 6.6$m^2$ per 1 kW solar PV, following KOPIA (2021).

[i] Following [46] C. Wilson, A. Grubler, N. Bauer, V. Krey, K. Riahi, Future capacity growth of energy technologies: are scenarios consistent with historical evidence?, Climatic Change 118(2) (2013) 381-395. https://dx.doi.org/10.1007/s10584-012-0618-y, we used the expansion rate as the change of installed capacity in a given time period (GW) over the total system size (GWh).

# 5. Conclusions

Korea's recent declaration of 2050 carbon neutrality is an important step to join the global efforts to limit climate change to 1.5°C above the pre-industrial level. With 30 years left to its carbon-neutrality target, we have assessed plausible carbon neutrality pathways while considering the current climate policies and major technology options. The challenge is unique as the energy system transition is being made for the currently carbon-intensive economy with the simultaneous phase-out of coal and nuclear power and the rapid introduction of clean technologies such as solar and wind. And the transition requires intensive land and geologic resources. Based on the GCAM integrated assessment framework, this paper has developed the first-of-a-kind long-term net-zero pathways of Korea's energy system, evaluating the feasibility of the 2050 carbon-neutrality target. It also analyzed the role of the power sector and its coupling with other sectors, and the value of critical mitigation technologies for the 2050 carbon- neutrality transition.

Our research findings support the following conclusions. First, overall, the carbon-neutrality target would require rapid decarbonization of the energy system over the next three decades and incur substantial policy costs, regardless of technology scenarios examined in our study.

Second, Korea's current policies would fall short of what its existing 2030 NDC target and the recently proposed revision would require (24.4% and 40% reductions relative to the 2018 level by 2030, respectively). The policy gap increases even further when emissions projected by the current policies are compared against the straight-line emissions path towards 2050 carbon neutrality.

Third, the power sector would play a pivotal role in Korea's carbon-neutrality transition. Across all scenarios, the power sector's rapid decarbonization is combined with the rapid electrification of the end-use sectors. Coal power phases out by the 2030s. Forth, and in the longer term, negative emissions technologies (NETs) come online in a considerable proportion to offset hard-to-abate emissions from the economy. This pattern is consistent across all of the net-zero scenarios explored in this paper.

Last, our research indicates that renewable power would continue to ramp up over the next three decades. However, its expansion rate is highly dependent upon the availability of nuclear power and CCS. Expansion of nuclear power would reduce the need for a rapid build-up of renewables and CCS. This alleviates the stress on terrestrial and geological systems. On the other hand, the combination of limited CCS and limited nuclear power would require unprecedentedly rapid ramp-up of solar and wind power. This could result in significantly higher policy costs, a measure of the difficulty of achieving the carbon neutrality target.

The scenario exercise and assessments performed in this paper offer two key insights into the development of climate and technology policies in Korea. First, strengthening of national climate policies is needed to achieve the ambitious carbon neutrality by 2050. The existing 2030 NDC target would need to be ratcheted up to send a credible signal to investors and stakeholders of the country's commitment to carbon neutrality. In this sense, the recently proposed upward revision of the NDC target from 24.4% to 40% is encouraging. For its implementation, the emissions cap for the Korean emissions trading scheme (K-ETS) would need to decline over time, with its long-term schedule announced early on in a way that is compatible with the carbon-neutrality target. For sectors currently not covered by the K-ETS, the introduction of an economy-wide carbon tax that rises over time in line with the K-ETS allowance price would help achieve carbon neutrality. As the transition is likely to provide substantial challenges to the Korean economy, revenues from auctioned emissions allowances and carbon taxes can be re-distributed to support the transition of presently carbon-intensive industries with limited option for decarbonization [47]. Such funds could also be directed toward small- and medium-sized enterprises that face difficulty in managing transition risks and workers seeking job transitions from fossil-fuel sectors [48].

Second, a 'technology-neutral' approach to 2050 carbon neutrality could help reduce the transition cost and increase the feasibility of the target. As our study indicates, exclusive reliance on renewable power without any contributions from nuclear and CCS technologies would require unprecedentedly rapid transition, while putting much stress on terrestrial and geological systems in Korea. Such rapid transition could also compromise electricity system reliability and economic security. The nuclear phase-out plan was originally introduced before the introduction of the carbon-neutrality goal. Facing the dual challenge of power system reliability and climate mitigation, it would be worth reconsidering the original nuclear phase-out plan. Any new nuclear power would need to be built with increased safety and waste disposal plans to enhance its social acceptance. Increased R&D support for carbon capture, utilization and storage (CCUS) technologies and exploration of $CO_2$ storage reservoirs would also help increase the viability of the carbon neutrality. $CO_2$ storage would provide hard-to-abate sectors with low carbon technology options.

In addition, the technology neutrality would serve the long-term planning for the 2050 carbon neutrality. Technology planning and R&D support should take into account that every ton of remaining positive emissions must be offset by an equal ton of negative emissions provided either by natural sinks or engineered $CO_2$ removal services. Just as there is a strong support for the $CO_2$ mitigation in the near term, there would also need to be equivalent support for longer-term negative emissions technologies (e.g., DAC, BECCS, and reforestation), which present high-risk high-return proposition for achieving carbon-neutrality.

Although the assessment presented in this paper provides a major improvement in our understanding of carbon neutrality transition, it is not without limitations. Here we highlight what we consider to be important areas for future research. First, although the scenarios presented in this paper represent the entire economy, they are modeled at a highly aggregated level. Most importantly, disaggregating the industry sector to account for technology and operational differences, and particularly those sensitive to climate policies would be useful. Second, the current analysis used exogenous pathways for low-carbon technologies, not exploring the endogenous impact of underlying climate policies on the rate of technological change and the emergence of radical mitigation technologies. Such a representation of endogenous technological change would be particularly relevant to enrich our understanding of the requirements and feasibility of the carbon neutrality target to be achieved within a short time frame.

Our research marks the first step towards developing feasible technology pathways towards carbon neutrality for Korea. The more scientific findings produced in this area, the better we can support policymakers and stakeholders to achieve this ambitious goal of carbon neutrality. A global community of collective effort can join forces, and together, the 1.5 goal can be achieved.

## Acknowledgements & Funding:

This publication was produced with the financial support of the European Union's Partnership Instrument. JE was also supported by the National Research Foundation (NRF) of Korea grants funded by the Korean government (NRF-2019K1A3A1A78112573). HM and CB were supported by the Global Technology Strategy Program. The authors would like to give special thanks to Joojin Kim and Kyungrak Kwon at Solutions for Our Climate for their support and feedback, especially in the early part of this project. The views and opinions expressed in this paper are those of the authors alone. They do not necessarily state or reflect those of the funding agencies and should not infer any official endorsement.

# Tables and Figures

Table 1. Definition of the scenarios

|  | CurPol | NZ2050 | NZ2050_Nuc | NZ2050_limCCS |
|---|---|---|---|---|
| New Nuclear Unit | Not allowed after 2024 | Not allowed after 2024 | Freely-competing based on technology costs | Not allowed after 2024 |
| CCS Storage Non- Energy Cost | 1,000 (2020USD/ tCO2) | 1,000 (2020USD/ tCO2) | 1,000 (2020USD/ tCO2) | 3,000 (2020USD/ tCO2) |
| Direct Air Capture Non-Energy Cost | 200 (2020USD/ tCO2) | 200 (2020USD/ tCO2) | 200 (2020USD/ tCO2) | 330 (2020USD/ tCO2) |
| Scenario description | Reflecting current and planned energy and climate policy measures | Assuming policy measures in CurPol and implementing additional climate policies to achieve net-zero GHG emissions by 2050 with the straight declining path | Implementing NZ2050 except that the expansion of nuclear power is allowed based on its economics | Implementing NZ2050 except that CCS deployment is restricted due to its assumed higher storage & capital costs of DAC |

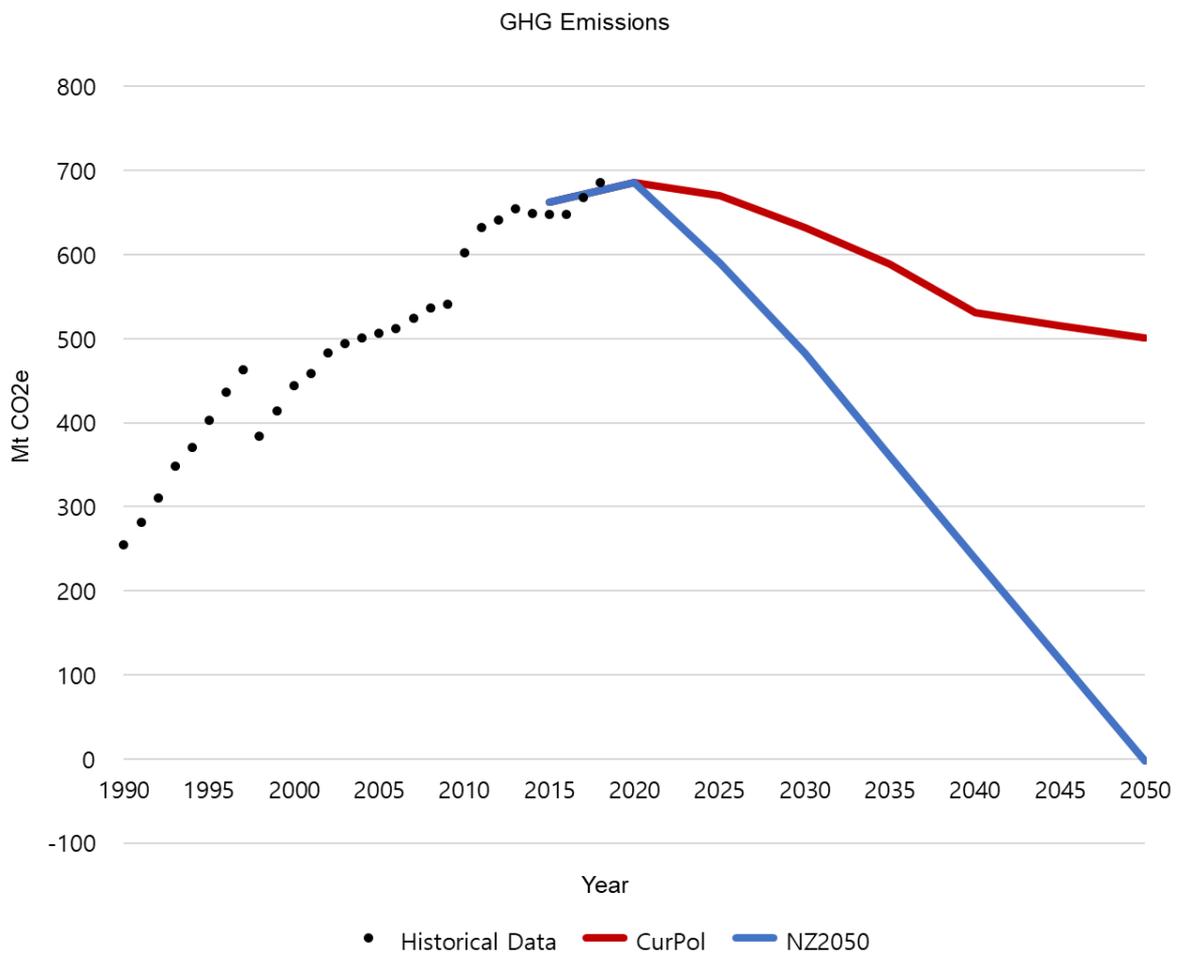

Figure 1. National total GHG emissions

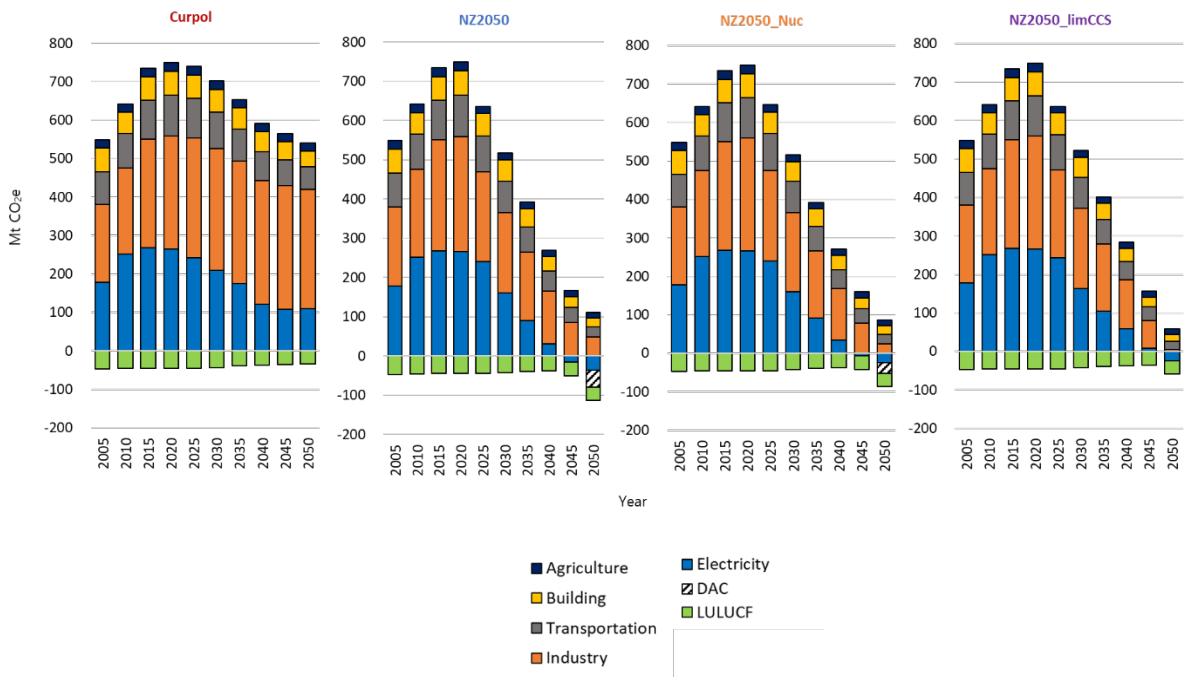

Figure 2. National GHG emission by sector

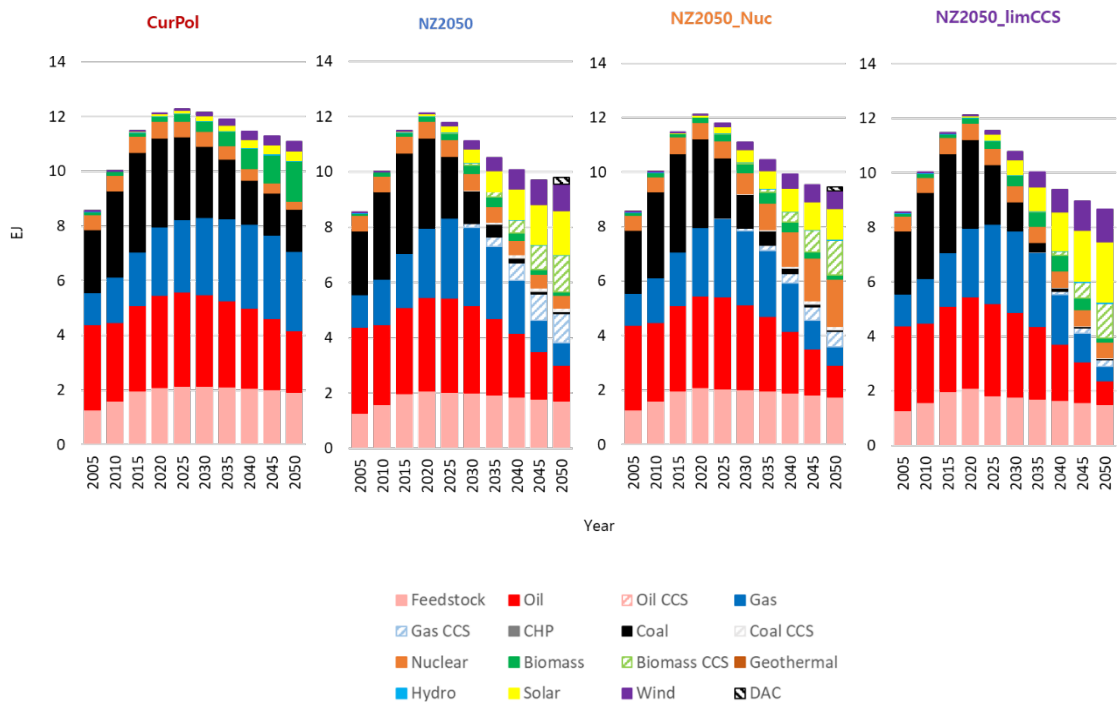

Figure 3. National primary energy consumption by fuel

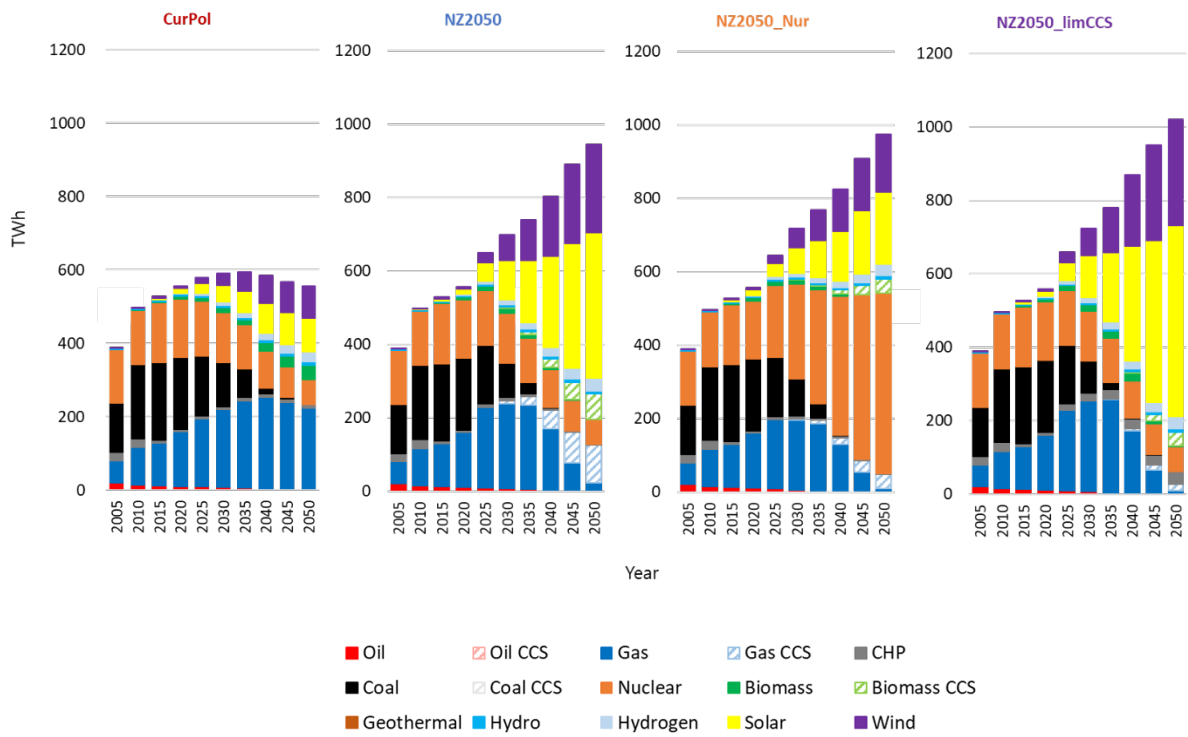

Figure 4. National electricity generation by technology

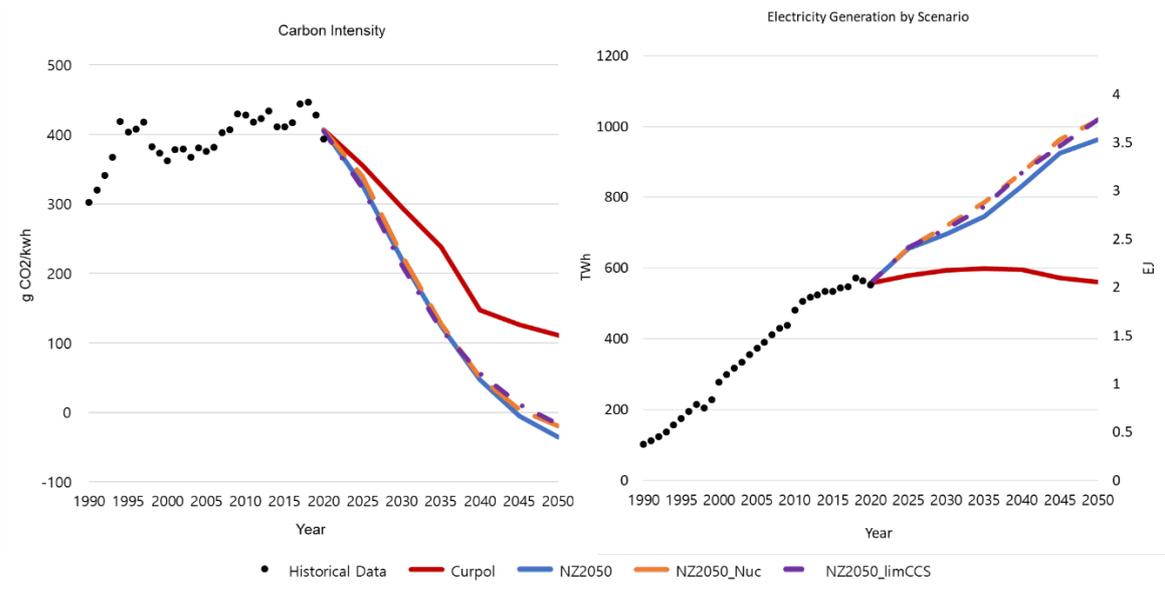

Figure 5. Carbon intensity of electricity generation (left) and total electricity generation (right) by scenario

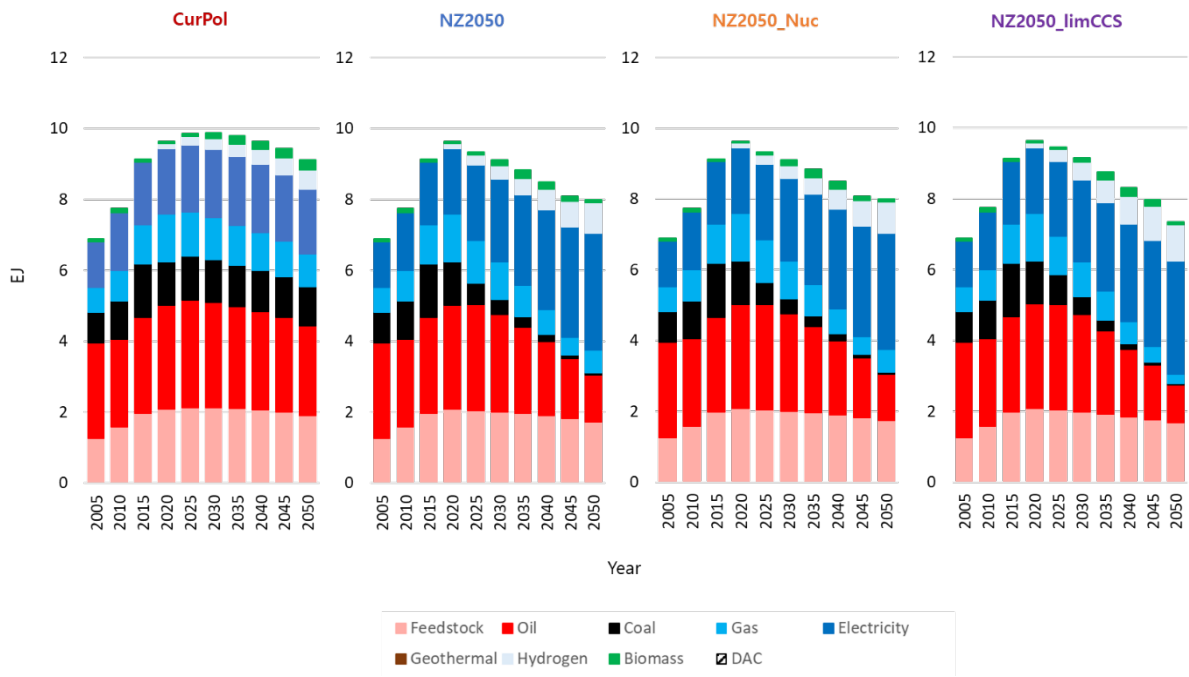

Figure 6. National final energy consumption by fuel

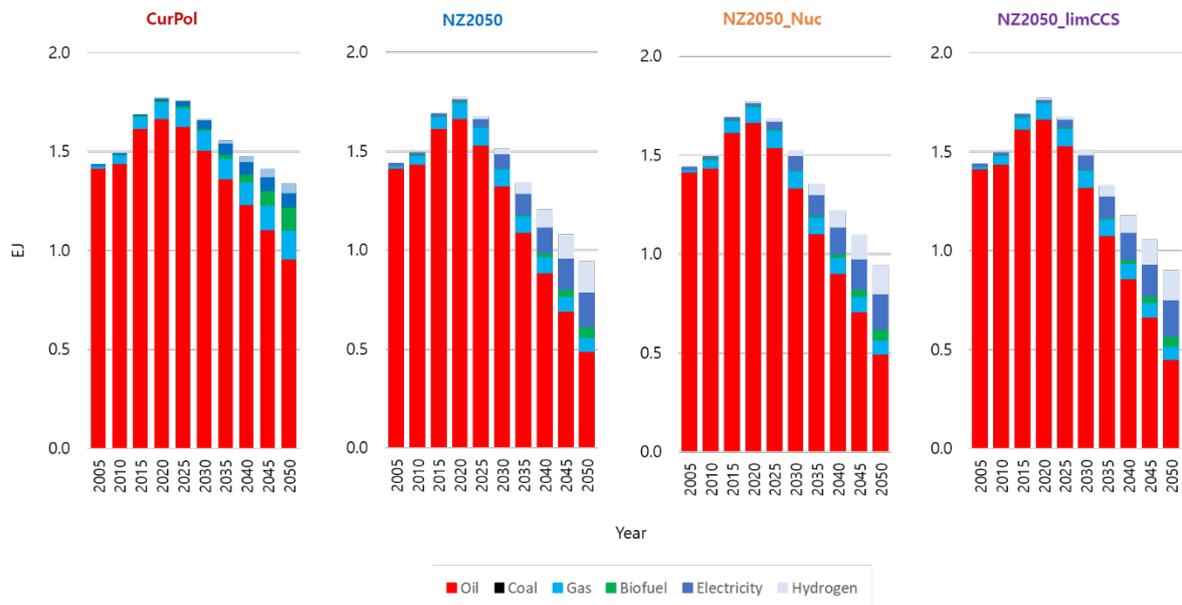

Figure 7. National transportation energy consumption by fuel

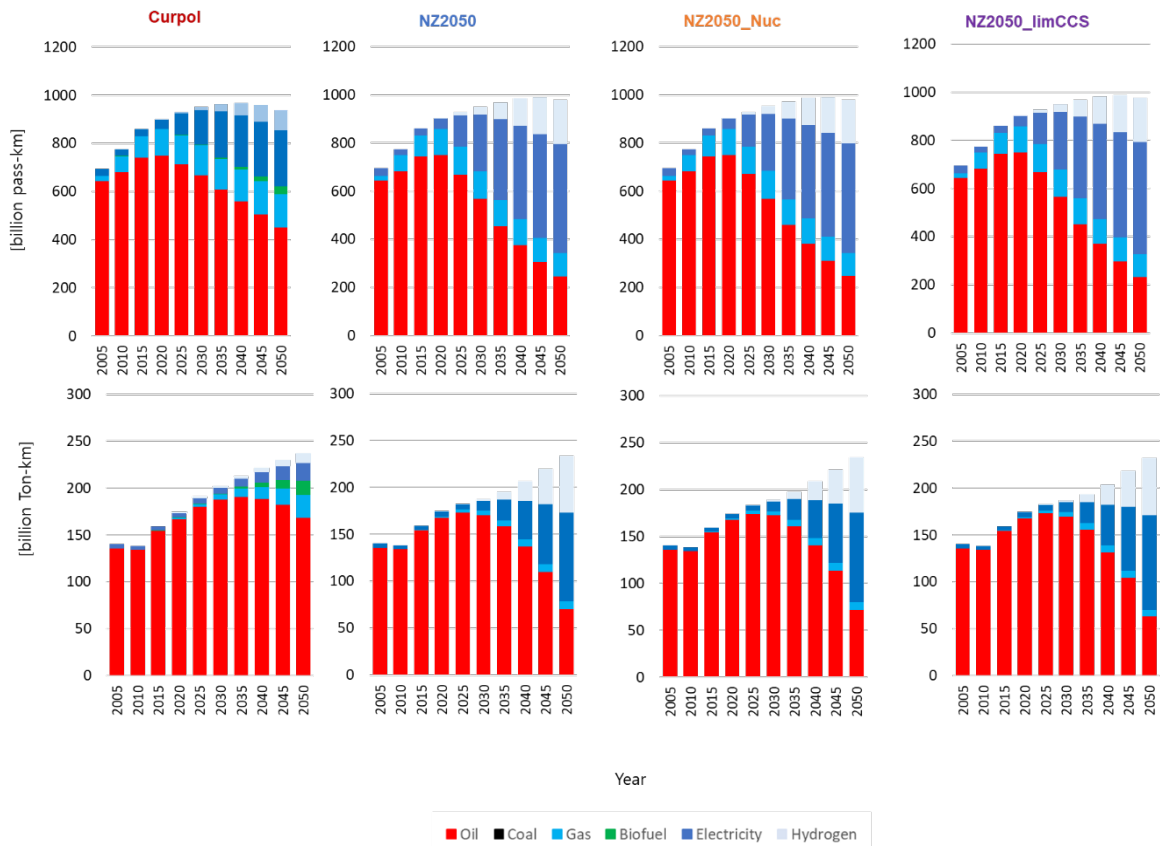

Figure 8. Passenger transport service (top) and freight transport service (bottom) by fuel

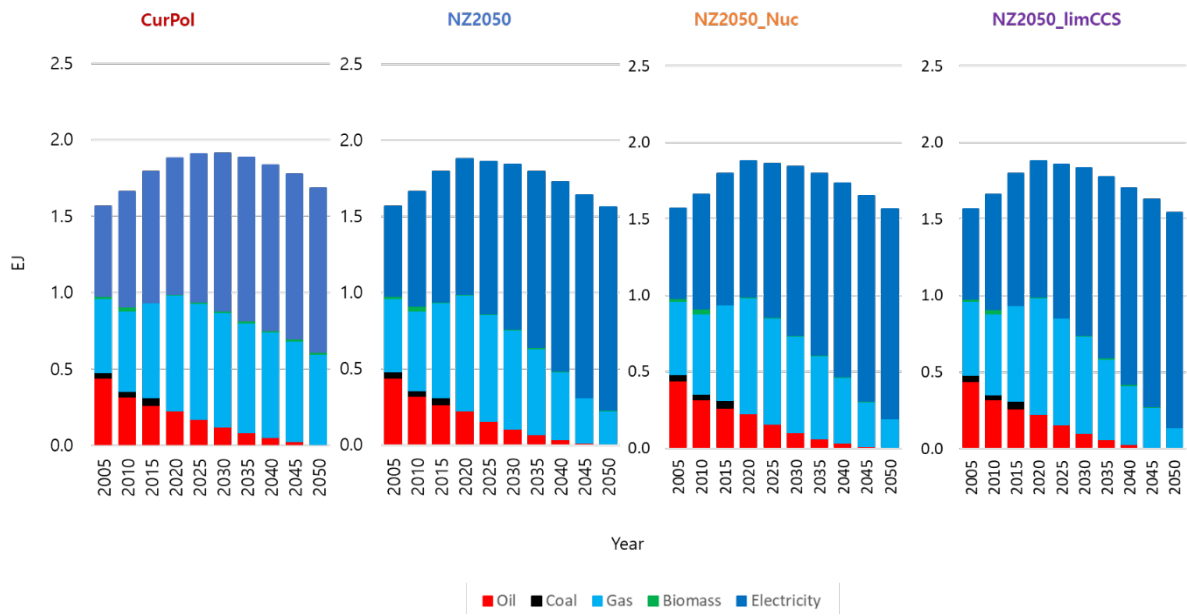

Figure 9. Buildings final energy consumption by fuel

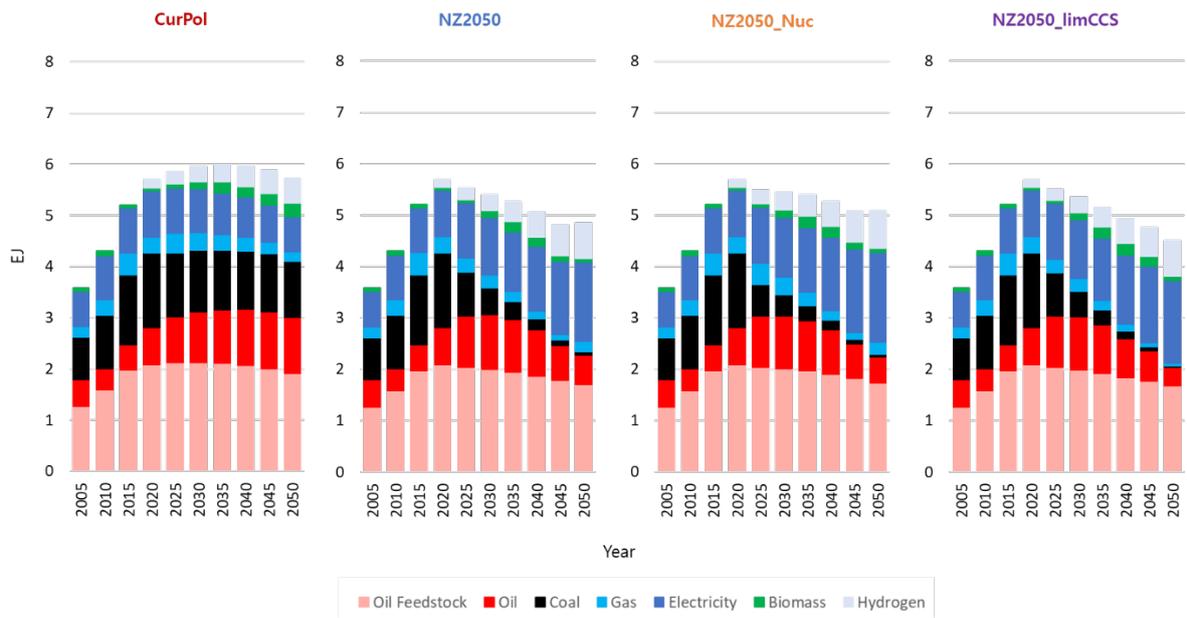

Figure 10. Industry final energy consumption by fuel

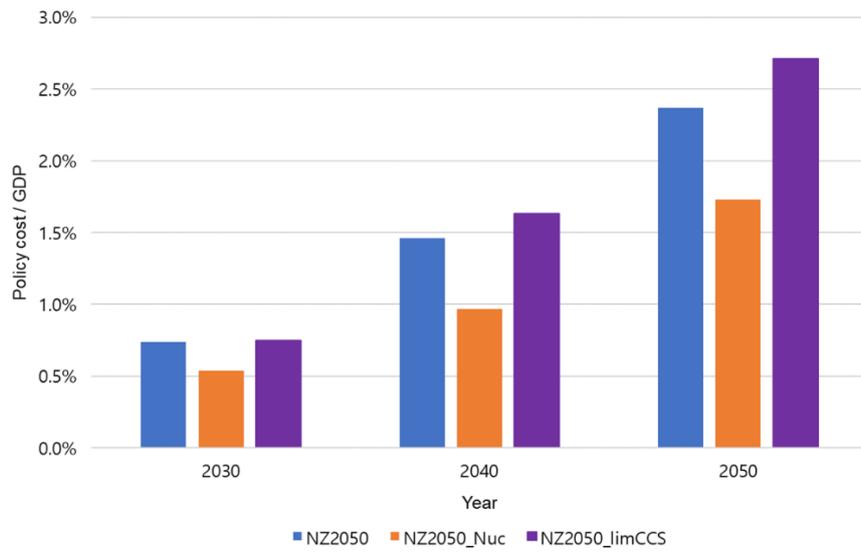

Figure 11. Policy costs over time

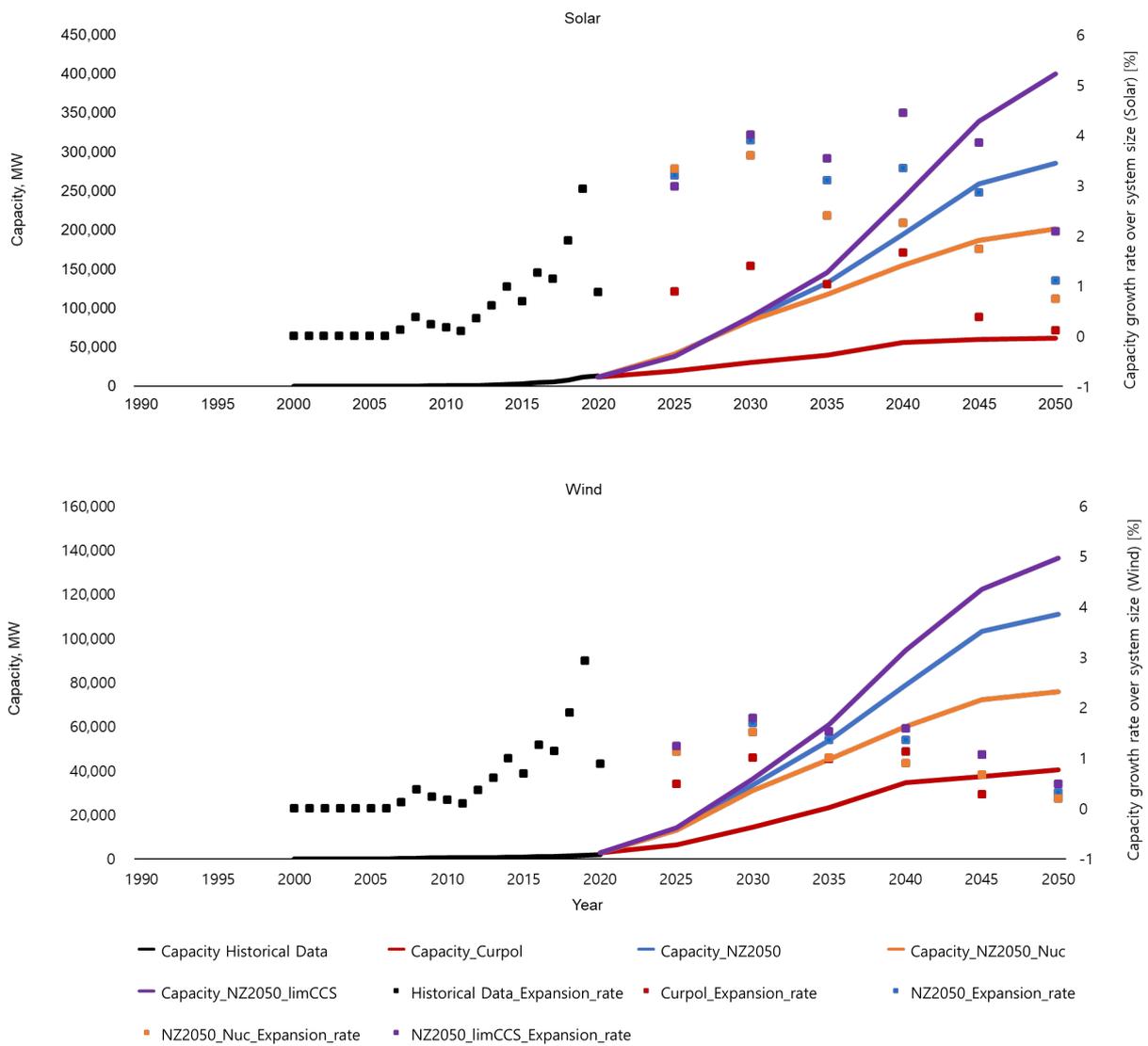

Figure 12. Power capacities and growth rates of solar (top) and wind (bottom) by scenario

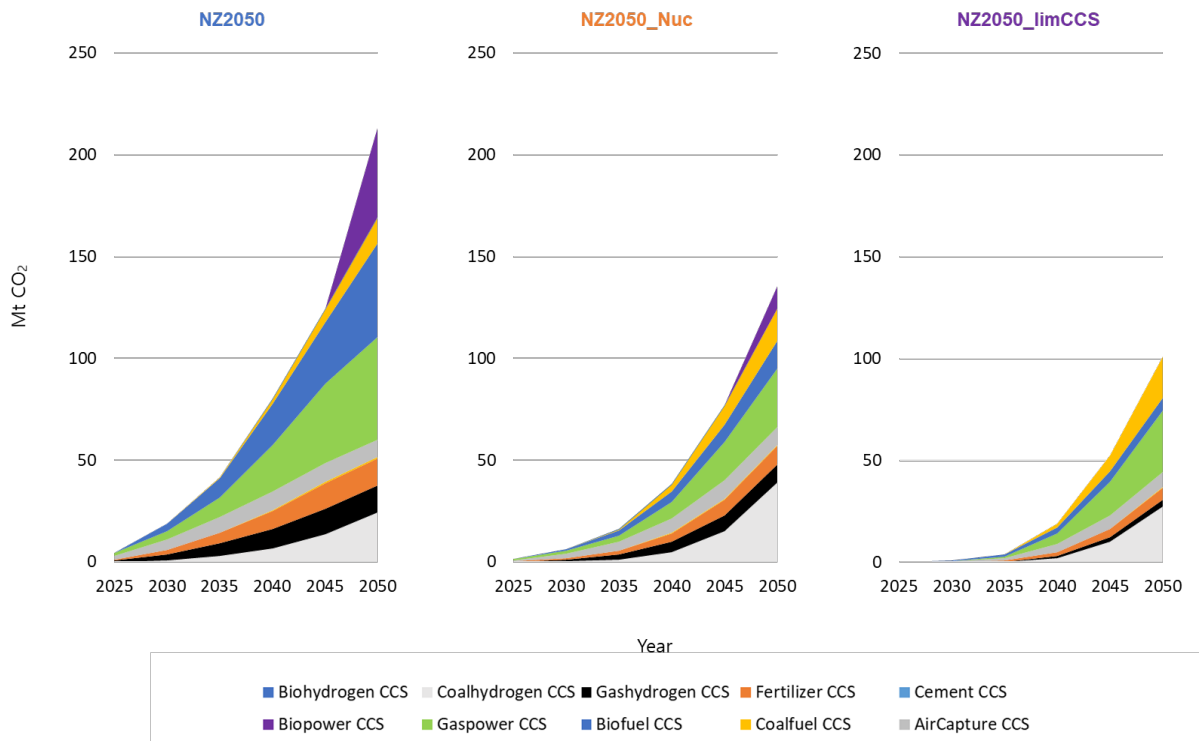

Figure 13. Annual CO$_2$ sequestration

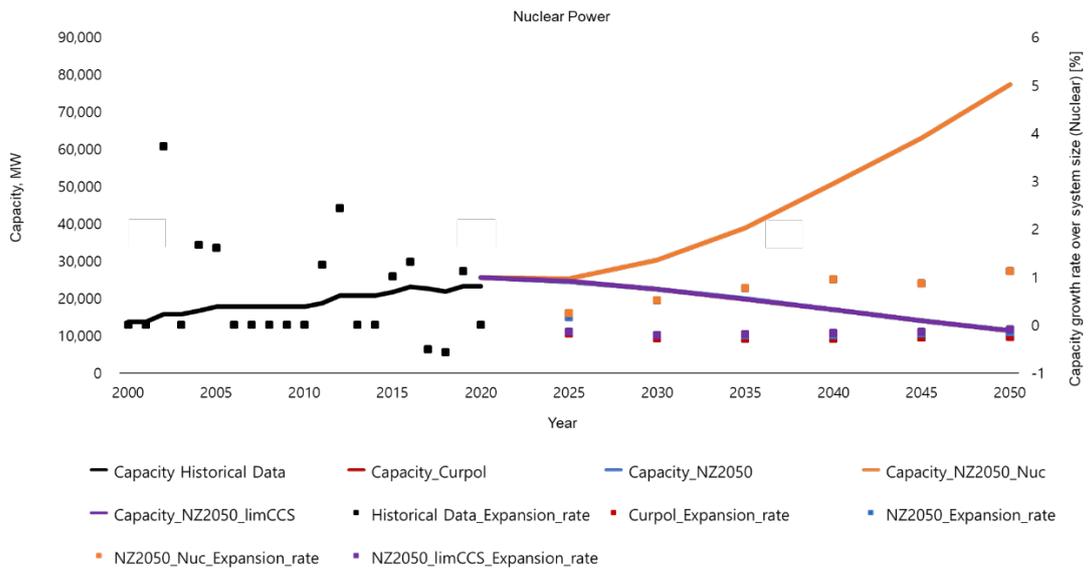

Figure 14. Nuclear power capacity and growth rate by scenario